\documentclass[a4paper,11pt]{article}
\usepackage[english]{babel}
\usepackage{a4wide}
\usepackage[latin1]{inputenc}
\usepackage{amsmath}
\usepackage{amsfonts}
\usepackage{amssymb}
\usepackage{epsfig}

\usepackage{graphics}
\usepackage{graphicx}
\usepackage{latexsym}
\usepackage{color}
\usepackage{marginnote}
\usepackage{url}
\usepackage{hyperref}

\begin{document}
	
	\title{Routes planning in a condominium company using mathematical optimization}
	
	\author{Lara Rocha \\ Coimbra Business School - ISCAC \\ Polytechnic University of Coimbra, Portugal \\
		e-mail: \texttt{a2019136971@alumni.iscac.pt}
		\and Pedro Martins \\ Coimbra Business School - ISCAC \\ Polytechnic University of Coimbra, Portugal, \\ Center for Mathematical Studies (CEMS.UL) \\ University of Lisbon, Portugal \\
		Centre for Organisational and Social Studies (CEOS.PP) \\ Technical University of Porto, Portugal \\
		e-mail: \texttt{pmartins@iscac.pt}}
	
	\maketitle 

	\begin{abstract}
		This work focuses on planning daily routes for a group of employees of a condominium company.
		
		Given a set of employees and a set of customers with pre-scheduled visits, we want to build doubly-open routes to guarantee visits to all customers, ensuring that each customer is visited by an employee with the appropriate skills to carry out the required activity. The travel times between customers and the service time in each customer are known. The problem includes time windows for visiting customers, each employee has a maximum working time, and the daily working period is between 9 AM and 6:30 PM. The routes respect the lunch break period if they extend from morning to afternoon, which means that lunch break can be avoided if the route is fully completed in the morning or in the afternoon. The aim is to minimize the distance covered by all routes with a penalty for including lunch breaks.
		
		The distances and times between customers are collected via an API to Microsoft Azure; and the solutions are then displayed on Bing Maps.
		
		The results obtained show the operational advantages of an efficient and effective vehicle routing planning in a specific condominium management company. The model can be extended to other service provider companies.

		
		
	\end{abstract}
	
\section{Introduction} \label{Sec1:Intro}

The construction of vehicle routes in pre-scheduled field service operations has long benefited from developments in mathematical optimization. This study addresses the problem of planning daily routes, with or without lunch breaks, for a team of collaborators working for a condominium management company.

In the context of condominium services, efficient routes planning is essential to ensure timely and consistent building maintenance, as well as to preserve a high level of residents' satisfaction. Ineffective planning can lead to unnecessary travel, unbalanced workloads, and increased operational costs. In contrast, routes optimization contributes to cost reduction, enhance collaborators productivity, and improve service quality, making the company more efficient and competitive in the long term.

Condominium management encompasses a wide range of service operations that are essential for the proper functioning of residential buildings. To meet service expectations, maintenance activities must be carried out efficiently and with minimum delay, ensuring a swift response to residents' needs. A recurring operational challenge in this sector involves organizing daily vehicle routes to fulfill scheduled visits and maintenance tasks. This is a regular challenge as companies need to serve multiple clients each day, each with specific service requirements and time constraints.

These routes are previously planned and do not include emergency services. When an emergency service is requested, the company uses collaborators exclusively for this purpose or triggers a re-optimization process of the routes, with possible damage to the service that has not yet been provided. Therefore, the problem under study in this paper will not consider the inclusion of emergency services. Instead, it focuses just on planning pre-scheduled requests.


Despite the operational importance of routes planning, many companies still rely on manual scheduling or, in some cases, do not plan them at all, addressing service requests as time and availability allows. Such practices undermine service consistency and organizational efficiency, which can compromise scalability and sustainable growth.

The problem considers: 
\begin{itemize}
	\item one route per collaborator;
	\item each collaborator can cover a predefined set of clients, according to the characteristics of the service required;
	\item all routes must be carried out during working hours, between 9 AM and 6:30 PM;
	\item each route has a time limit in accordance with the collaborator's availability;
	\item each route starts on the first client and ends after serving the last, so it can be considered a doubly-open route;
	\item each client has a time window for the start of the visit; and
	\item a lunch break is considered if the route crosses the window between 12:30 AM and 2 PM.
\end{itemize}

The objective of the problem is to find feasible routes for the collaborators, minimizing the sum of the total distance traveled, with a penalty on the inclusion of a lunch break in each route.

It is important to justify the relevance of the conditions previously stated for companies providing condominium services. In effect, given the technical specificity of certain operations requested by the clients and the collaborators' skills, it is important to previously classify the clients that each collaborator is able to serve. This way, each collaborator will be able to cover only a subset of the entire set of clients in the instance under analysis. In addition, we are not interested in the vehicles' capacities but on the capability of each collaborator to make his/her scheduled route, within the company's time working period and the collaborators' time restrictions. Furthermore, the daily working period of these companies is limited, excluding emergency services. At the same time, the working period of each collaborator is also fixed, depending on his/her own availability or contract restrictions. In addition, the routes are doubly open, which means that the collaborators use their own vehicle or a company vehicle that is also used for private life. In these cases, each collaborator starts its journey at the first client and ends at the last one, that is, the collaborator can come directly from home to the first client and go directly home from the last client, with no need to pass through the office. This use of a vehicle for work and personal life purposes is very common in service companies. This doubly open feature of the routes allows us to see the problem as a partition of the set of clients into Hamiltonian paths (ignoring lunch breaks).

Another aspect to consider involves the use of time windows on each client. In effect, the possibility of clients being able to set time windows allows them to better organize their personal lives; but greatly limits the scheduling of visits by the company when building routes for its collaborators. This means that the use of time windows can be very beneficial for clients, but very damaging to the company collaborators' working time, making the process of building vehicle routes even more challenging. Despite these limitations, the use of time windows is an important aspect that many condominium companies seek to adopt, doing so to provide a better service and gain a competitive advantage over competing companies.

Finally, the inclusion of a lunch break is also unavoidable when the routes extend over long periods, covering morning and afternoon. This is an aspect protected by labor laws and is also present in the problem under study.
Furthermore, in our case, the routes must respect a specific lunch break period that is not counted in the collaborator's total route time. Similarly, the distance traveled between the two clients that precede and follow lunch break is also not counted. For these reasons, preference will be given to routes that can avoid the use of lunch breaks, which is modeled by penalizing, in the objective function, the inclusion of lunch breaks in the routes under construction.

\medskip
The required data in each instance includes the distances and travel times between clients, the expected service duration at each client, the time windows within which each client must be visited, the sets of clients that can be covered by each collaborator, and the daily working period of each collaborator. Distances and travel times between clients are collected using a Microsoft Azure API, and solutions are visualized through Bing Maps.

The results obtained show the operational benefits of implementing efficient and effective vehicle routing strategies for field service operations within a condominium management company.

The main contributions of this paper involve the discussion of condominium daily routing planning for field service operations from an applied point of view. It also discusses the inclusion of doubly-open routes construction and further contributes to the modeling discussion on the inclusion of lunch breaks in the routes, bringing the possibility of not including a lunch break if the route is carried out entirely in the morning or entirely in the afternoon. To the best of our knowledge, these aspects have rarely been discussed in the literature. In addition, our model uses fewer binary variables than other models in the literature that address vehicle routing problems with lunch breaks in the routes.

In the next section, a literature review on related problems is discussed. Then, in Section \ref{Sec3:Form}, a mixed integer linear programming formulation is proposed to the problem in hands, including ways to strengthen the associated linear programming relaxation. Computational tests are conducted in Section \ref{Sec4:Tests} involving synthetic and real-world instances, to test the proposed model and to emphasize the practical relevance of the problem under study. The paper ends with a conclusions section.

\section{Literature review} \label{Sec2:Literature}

The study of Vehicle Routing Problems (VRP) has sought to address numerous practical aspects arising from the diverse operational environments in which the VRP has been applied. Some of these aspects are also brought to the problem under study, while others are less frequently encountered. This review begins by focusing on the most commonly studied characteristics (see, e.g., \cite{ZhaGeYanTon2022}, \cite{KonGayKec2022}).

Routes of services, namely field service operations involve performing maintenance, installation, or support activities directly at customer locations. Efficient planning of service routes is critical to reduce travel costs, minimize response times, and maximize resource utilization, including specialized technicians and vehicles. Given the complexity and variability of tasks, these operations are often modeled as variants of the VRP, enabling the optimization of both routes and resource allocation. Therefore, efficient organization of these routes is a central aspect of field service management, reflecting the need to balance logistic efficiency with the quality of service provided to customers (\cite{BlaArgCaoHalKno2003}, \cite{FriBer2021}).

In many practical vehicle routing applications, operations are constrained not only by the limited number of available vehicles but also by fleet heterogeneity and the individual working times of operators. These conditions correspond to a variant of the VRP known as the Fixed Fleet Heterogeneous VRP, where each vehicle has distinct capacities and characteristics and is associated with workers with varying service times (\cite{AhmedYou2023}, \cite{LaCoeRe2018}). The limitation in fleet size and the diversity of the fleet significantly increase planning complexity, requiring optimization strategies that balance operational efficiency, adherence to time windows, and human resource constraints. Recent studies show that hybrid or metaheuristic approaches can be particularly effective in addressing these constraints, providing feasible and near-optimal solutions in scenarios with fixed and heterogeneous fleets (\cite{EshEsmMid2020}).

The Open Vehicle Routing Problem (OVRP) is a variant of the classic VRP that offers greater flexibility in route definition, allowing vehicles to start and finish their journeys in ways that differ from closed routes, in which the vehicle must start and return to the depot after serving all customers. In the OVRP, the vehicle may start its route at the first customer and return to the depot after completing all deliveries, or it may start at the depot and finish its journey after serving the last customer, without the need to return to the starting point. This feature makes the problem particularly relevant in contexts where logistics involve private vehicles or shared resources, such as deliveries carried out by employees using their own vehicles or when company vehicles are also used for personal purposes.
The main advantage of the OVRP over the traditional closed route model lies in the flexibility in route planning, which allows for a significant reduction in total costs and a more efficient use of resources. This increase in flexibility translates into better adaptation to transportation contexts in which returning to the depot is not always required, allowing route optimization based on the specific needs of each company (\cite{Bra2004}, \cite{LetLysEgl2007}, \cite{LiGolWas}).
\cite{CabCorMen2022} developed a metaheuristic for the Doubly Open Park-and-Loop Routing Problem, a variant in which routes may start and end at different locations, without the need to begin or finish at the depot (doubly open routes), and which incorporates walking sub routes performed after the vehicle is parked. This study represents one of the most recent contributions and, to date, the only one identified being dedicated to doubly open routing models.

In most real-world applications of the Vehicle Routing Problem (VRP), the inclusion of temporal constraints is essential, as these directly influence the real adequacy of the routes. Among these constraints, the limitations associated with drivers' working hours are particularly relevant, often imposed by legal regulations aimed at ensuring road safety and compliance with labour laws. Such regulations define, for example, the maximum allowable daily and weekly driving time, as well as mandatory rest periods.
These requirements are modeled in specific VRP variants, such as the Vehicle Routing Problem with Drivers' Working Hours (VRPDWH), which extends the classical formulation by incorporating limits on driving time and mandatory breaks \cite{GoeGru2006}. In addition to labor-related restrictions, it is also common to impose a limit on the total route duration, including both travel time and service time, to ensure that each route does not exceed a predefined maximum value \cite{CokSubLarStins2023}. Although necessary, these constraints increase the complexity of the problem, as the integration of breaks and temporal limits cannot compromise service levels or the fulfilment of delivery deadlines. They also bring complexity because these conditions act like knapsack constraints in the VRP model, making its resolution harder to reach.

The Vehicle Routing Problem with Time Windows (VRPTW) is also a well-known extension of the VRP in which each customer must be served within a fixed time window, and any violation renders the solution infeasible. Variants such as the Vehicle Routing Problem with Soft Time Windows (VRPSTW) and the Vehicle Routing Problem with Hard Time Windows (VRPHTW) introduce flexibility: in the VRPSTW, serving a customer outside the time window incurs a financial penalty, while in the VRPHTW, service must occur within the time window, although the vehicle is allowed to arrive early and wait until the start of the service \cite{ElsAwad2020}, \cite{Dumasetal1991}. In many cases, vehicle routing problems with time windows are enriched with other attributes to better answer and capture real world characterisitcs \cite{ArmBatPerBri2015}.

Some papers in the literature discuss the inclusion of a lunch break in each route, being used if the worker's labor period encompasses morning and afternoon, namely in (\cite{CoeGagRenRui2016}, \cite{Baz2025}, \cite{Khaetal2025}). The work of Bazirha (\cite{Baz2025}) proposes a new characterization of lunch breaks within a VRPTW formulation that minimizes the total travel time, comparing their model characterization with those proposed in \cite{BurLarRop2012} and \cite{CoeGagRenRui2016}. In \cite{XiaDriHas2018}, the authors consider flexible lunch break requirements, allowing the resolution to decide whether to include or not a lunch break in each route. Our model follows a similar flexible methodology. The modeling characterization proposed in \cite{CoeGagRenRui2016} considers lunch breaks as dummy nodes and sets the lunch period break as a time window as for the other nodes, which means that they apply time windows to all nodes, including lunch break dummy nodes. This methodology is also folloed in our paper.

The need of a lunch break is also discussed in routing problems that address home healthcare workers scheduling (see, e.g., \cite{LiuYuaJia2017}). However, a less common aspect involves the possibility of deactivating the lunch break if the route operates exclusively in the morning or exclusively in the afternoon. This feature is also discussed in the present paper.

In all the previously mentioned approaches addressing lunch breaks in the routes, lunch pause is compulsory, regardless of the duration of the routes. This is a natural feature when collaborators take a full-day work period only performing service operations in the field. In our case, the collaborators may carry out other activities, becoming more productive if having an entire half-day for those other duties.

Heuristic methodologies for routing problems with lunch breaks are countless. Many of the works reported above and below resort to those techniques to answer routing problems, especially when exact methodologies fail to provide a response in realistic amount of time. The present paper only discusses modeling approaches to the condominium field service routing problem here considered, with no insights on heuristic methodologies.

From an applied point of view, routes planning is becoming increasingly common and important for companies that rely on road transportation. There are countless practical applications in this area, such as planning routes for school buses, delivering and collecting parcels, delivering and collecting money from banks or automated teller machines (ATM), collecting garbage, providing home services, home clinical services
(nursing care, post-surgery care, outpatient care), maintenance operations, among others (see, e.g.,\cite{BurLarRop2012}, \cite{LiuYuaJia2017}, \cite{FarLys2021}).

\section{Mathematical formulation} \label{Sec3:Form}

Considering the problem described above, this chapter presents a mathematical formulation for modeling it. The proposed model is a Mixed Integer Linear Programming formulation that incorporates a set of additional flow variables in order to construct flow conservation constraints and flow limits. These constraints allow to prevent the existence of subcircuits in the solutions. The use of subcircuit elimination conditions by means of flow variables is a well-known strategy to build a formulation with a polynomial number of variables and constraints for the vehicle routing problem being addressed (\cite{GavGra1978}). Furthermore, additional constraints will be considered in order to strengthen the associated linear programming relaxation, whose results are reflected in the computational tests discussed in Chapter \ref{Sec4:Tests}.

Before introducing the model, we start by defining the parameters and variables. Thus, parameter $n$ represents the number of clients to visit on a given day; while parameter $m$ represents the number of collaborators available, where $M$ is the set of all collaborators. In addition, $k$ is the index that will be used to represent the collaborators, that is, $k \in M$.

The vehicle routing problem under analysis is characterized on graph $G_0=(V_0,A_0)$, where the sets of nodes and arcs are defined below, including specific subsets that will be useful for constructing the model.

Sets and subsets of nodes:
\begin{itemize}
	\item $V1 = \{1,...,n\}$, is the set of clients to visit;
	\item $V2 = \{n+1,...,n+m\}$, is the set of lunch nodes that act as fictitious clients, where node $n+k$ represents the lunch node of collaborator $k$, for all $k \in M$;
	\item $V = V1 \cup V2$, is the set of all clients, excluding the depot ($0$), but including the fictitious clients (lunch nodes);
	\item $V_0 = \{0\} \cup V$, is the set of all clients, including the depot (node $0$), the clients to visit ($V1$) and the fictitious clients ($V2$) that represent lunch breaks;
	\item $V1^k \subseteq V_1$, for all $k \in M$, is the set of clients that collaborator $k$ can cover;
	\item $V^k = V1^k \cup \{n+k\}$, for all $k \in M$, is the set of clients that collaborator $k$ can visit, including the respective fictitious client;
	\item $V_0^k = V^K \cup \{0\}$, for all $k \in M$, is the set of clients that collaborator $k$ can cover, including the depot and the respective fictitious client.
\end{itemize}

Sets and subsets of arcs:
\begin{itemize}
	\item $A1 = \{(i,j) \in V1 \times V1 : i \neq j\}$, representing the set of arcs among clients to visit, that is, it includes all arcs in the graph excluding those incident on the depot and those incident on the lunch nodes;
	\item $A2 = \{(i,j) \in V2 \times V2 : i \neq j\} \cup \{(0,j) : j \in V2 \} \cup \{(i,0) : i \in V2 \}$, representing the set of arcs among lunch nodes and among the depot and lunch nodes (these arcs do not belong to $A_0$));
	\item $A_0 = \{(i,j) \in V_0 \times V_0 : i \neq j\} \setminus A2$, representing all connections among the clients, the depot and the lunch nodes, excluding the arcs in $A2$, namely the arcs between the depot and the lunch nodes and the arcs among lunch nodes;
	\item $A = A_0 \setminus \left( \{(0,j) \in A_0 : j \in V\} \cup \{(i,0) \in A_0 : i \in V\} \right)$, representing the set of arcs in the graph, excluding those incident on the depot;
	\item $A1^k = (\{(i,j) \in A1: i,j \in V1^k \text{ and } i \neq j\}) $, representing the set of arcs that collaborator $k$ can traverse, just among clients that collaborator $k$ can visit;
	\item $A^k = \{(i,j) \in A: i,j \in V^k \text{ and } i \neq j\ \}$, representing the set of arcs that collaborator $k$ can traverse, excluding arcs incident on the depot;
	\item $A_0^k = (\{(i,j) \in A_0: i,j \in V_0^k \text{ and } i\neq j\ \})$, representing the set of arcs that collaborator $k$ can traverse, including the depot and the fictitious client, but excluding connections between the depot and the fictitious client.
\end{itemize}

Sets and subsets of collaborators:
\begin{itemize}
	\item $M_j \equiv$ representing the set of collaborators that can serve client $j$, for all $j \in V1$;
	\item $M_{ij} \equiv$ representing the set of collaborators that can serve both nodes $i$ and $j$, for all $i,j \in V_0$.
\end{itemize}

These sets are used to assist in the definition of the parameters and variables; and then in the model.

List of parameters:
\begin{itemize}
	\item $d_{ij} \equiv$ (route) distance from node $i$ to node $j$, for all $(i,j) \in A1$;
	\item $t_{ij} \equiv$ (route) traveling time from node $i$ to node $j$, for all $(i,j) \in A1$;
	\item $s_i \equiv$ service time at node $i$, for all $i \in V$;
	\item $T^k \equiv$ maximum daily labor time for collaborator $k$, for all $k \in M$;
	\item $n^k \equiv$ total number of clients that collaborator $k$ is skilled to serve, for all $k \in M$, that is, $n^k = |V^k|$;
	\item $a_j' \equiv$ starting instant of the time window on node $j$, for all $j \in V$;
	\item $a_j'' \equiv$ ending instant of the time window on node $j$, for all $j \in V$;
	\item $H \equiv$ total daily operating time of the company;
	\item $L1 \equiv$ starting time for the lunch break;
	\item $L2 \equiv$ ending time for the lunch break.
\end{itemize}

To simplify the construction and indexing of the formulation, it is assumed that the service time at the depot is assumed to be zero ($s_0 = 0$).

List of variables:
\begin{list}{ }
	\item $x_{ij}^k = \left\{
	\begin{tabular}{ll}
		1, & if collaborator's route $k$ goes through arc $(i,j)$ \\
		0, & otherwise
	\end{tabular}
	\right., \;$ for all $(i,j) \in A_0^k$ and all $k \in M$	
	\item $y_{ij}^k \equiv$ flow passing through arc $(i,j)$, corresponding to the number of clients already visited when collaborator $k$ travels through arc $(i,j)$, eventually including the lunch node, for all $(i,j) \in A_0^k$ and all $k \in M$
	\item $a_j \equiv$ instant time (in minutes) at which client $j$ is visited, for all $j \in V$
	\item $w^k = \left\{
	\begin{tabular}{ll}
		1, & if collaborator's route $k$ includes the lunch break \\
		0, & otherwise
	\end{tabular}
	\right., \;$ for all $k \in M$	
	\item $b1^k = \left\{
	\begin{tabular}{ll}
		1, & if all the service of collaborator $k$ is entirely in the morning \\
		0, & otherwise
	\end{tabular}
	\right., \;$ for all $k \in M$	
	\item $b2^k = \left\{
	\begin{tabular}{ll}
		1, & if all the service of collaborator $k$ is entirely in the afternoon \\
		0, & otherwise
	\end{tabular}
	\right., \;$ for all $k \in M$	
\end{list}

These sets of parameters and variables are used to build the initial flow formulation for the routing planning problem under discussion. This initial formulation is denoted by F0.

\begin{eqnarray}
	\medskip
	& \mbox{min} & \displaystyle\sum_{k \in M} \sum_{(i,j) \in A1^k} d_{ij} x_{ij}^k + \displaystyle\sum_{k \in M} \max_{(i,j) \in A1} \{d_{ij}\} w^k \label{eqn:f1} \\
	\medskip
	& \mbox{s.t.} & \displaystyle\sum_{k \in M_j} \sum_{(i,j) \in A_0^k} x_{ij}^k = 1\;, \;\;\forall j \in V1 \label{eqn:f2} \\
	\medskip
	& & \displaystyle\sum_{(i,n+k) \in A_0^k} x_{i,n+k}^k = w^k\;, \;\;\forall k \in M \label{eqn:f3} \\
	\medskip
	& & \displaystyle\sum_{j:(i,j) \in A_0^k} x_{ij}^k - \displaystyle\sum_{j:(j,i) \in A_0^k} x_{ji}^k = 0\;, \;\;\forall i \in V^k \mbox{ and } \forall k \in M \label{eqn:f4} \\
	\medskip
	& & \displaystyle\sum_{j \in V1^k} x_{0j}^k \leq 1\;, \;\;\forall k \in M \label{eqn:f5} \\
	\medskip
	& & - \displaystyle\sum_{i:(i,j) \in A^k} y_{ij}^k + \left( \displaystyle\sum_{i:(j,i) \in A^k} y_{ji}^k \right) + y_{j0}^k = \displaystyle\sum_{i:(j,i) \in A_0^k} x_{ji}^k\;, \;\;\forall j \in V^k \mbox{ and } \forall k \in M \label{eqn:f6} \\
	\medskip
	& & y_{ij}^k \leq (n^k - 1)x_{ij}^k \;, \;\;\forall (i,j) \in A^k \mbox{ and } \forall k \in M \label{eqn:f7a} \\
	\medskip
	& & y_{i0}^k \leq n^k x_{i0}^k \;, \;\;\forall i \in V1^k \mbox{ and } \forall k \in M \label{eqn:f7b} \\
	\medskip
	& & \displaystyle\sum_{(i,j) \in A_0^k \land (i \neq n+k)} s_i x_{ij}^k + \displaystyle\sum_{(i,j) \in A1^k} t_{ij} x_{ij}^k \leq T^k\;, \;\;\forall k \in M \label{eqn:f9} \\
	\medskip
	& & a_j + s_j \leq H \;, \;\;\forall j \in V1 \label{eqn:f10} \\
	\medskip
	& & a_j' \leq a_j \leq a_j'' \;, \;\;\forall j \in V \label{eqn:f11} \\
	\medskip
	& & a_j \geq (a_i + s_i + t_{ij}) - (1 - x_{ij}^k)H \;, \;\;\forall (i,j) \in A^k \;, \;\; i \neq j \mbox{ and } \forall k \in M \label{eqn:f12} \\
	\medskip
	& & \displaystyle\sum_{i \in V1^k} x_{i,n+k}^k + \displaystyle\sum_{j \in V1^k} x_{n+k,j}^k \leq 2w^k\;, \;\;\forall k \in M \label{eqn:f13} \\
	\medskip
	& & a_j + s_j \leq L1 + T^k(1 - b1_k) + T^k \left( 1 - \displaystyle\sum_{i \in V_0^k} x_{ij}^k \right) \;, \;\;\forall j \in V1^k \mbox{ and } \forall k \in M \label{eqn:f14a} \\
	\medskip
	& & a_j \geq L2 - T^k(1 - b2_k) - T^k \left( 1 - \displaystyle\sum_{i \in V_0^k} x_{ij}^k \right) \;, \;\;\forall j \in V1^k \mbox{ and } \forall k \in M \label{eqn:f14b} \\
	\medskip
	& & w^k = 1 - b1^k - b2^k \;, \;\;\forall k \in M \label{eqn:f15} \\
	\medskip
	& & x_{ij}^k \in \{0,1\}\;, \;\;\forall (i,j) \in A_0^k \mbox{ and } \forall k \in M \label{eqn:f17a} \\
	\medskip
	& & y_{ij}^k \geq 0\;, \;\;\forall (i,j) \in A_0^k \setminus \{(0,j): j \in V1^k \} \mbox{ and } \forall k \in M \label{eqn:f17b} \\
	\medskip
	& & a_j \geq 0\;, \;\;\forall j \in V_0 \label{eqn:f17c} \\
	\medskip
	& & w^k, b1^k, b2^k \in \{0,1\}\;, \;\;\forall k \in M \label{eqn:f17d}
\end{eqnarray}

The objective function (\ref{eqn:f1}) contains two components. The first one involves the minimization of the sum of the traveling distances of all routes, while the second one penalizes the inclusion of the lunch break in each route. As mentioned above, we are looking for solutions that avoid lunch breaks, forcing the operations to be entirely performed in the morning or in the afternoon, when possible. 

The set of constraints in the model ensures the coherence of the routes of the collaborators. Constraints (\ref{eqn:f2}) guarantee that each customer is visited exactly once. Constraints (\ref{eqn:f3}) activate the lunch-break variable whenever the route includes the corresponding fictitious node.
Constraints (\ref{eqn:f4}) force the route to be done by the same employee, imposing that an employee's arrival at a client implies that the same employee must also depart from that client. Constraints (\ref{eqn:f5}) ensures that each collaborator starts at most one route. The set of constraints ((\ref{eqn:f6}), (\ref{eqn:f7a}) and (\ref{eqn:f7b})), based on flow variables, eliminate sub-tours. They also link the two sets of variables $x$ and $y$, forcing the presence of arc $(i,j)$ every time an arc has flow and closing the flow in any arc not in the solution.
Constraints (\ref{eqn:f9}) limit the total route time according to each employee's contract or availability, and (\ref{eqn:f10}) enforce compliance with the company's operating hours. Clients' time windows are ensured by constraints (\ref{eqn:f11}), while (\ref{eqn:f12}) guarantee temporal consistency between visits, allowing waiting periods whenever necessary.
Constraints (\ref{eqn:f13}) link the inclusion of the lunch break to the edges incident to the fictitious node. Inequalities (\ref{eqn:f14a}) and (\ref{eqn:f14b}) characterize the cases when the collaborator works just in the morning or just in the afternoon, respectively, setting $L1$ to 1 if all clients are served in the morning; and sets $L2$ to 1 if the service in all start in the afternoon, respectively. Constraints (\ref{eqn:f15}) guaranty exclusivity between working the full day (with lunch) or working just in one of the shifts.
Finally, constraints (\ref{eqn:f17a}, \ref{eqn:f17b}, \ref{eqn:f17c}, \ref{eqn:f17d}) are domain conditions, defining the domain of the variables.

\vspace{0.5cm}

The linear programming relaxation version of F0 can be strengthened by adding an additional set of linking constraints (\ref{eqn:f20}).

\begin{equation}\label{eqn:f20}
	x_{ij}^k \leq y_{ij}^k\;, \;\;\forall (i,j) \in A_0\setminus \{(0,j): j \in V1^k \} \mbox{ and } \forall k \in M
\end{equation}

Inequalities (\ref{eqn:f20}) inhibit the presence of arc $(i,j)$ in the solution if there is no flow through it.

We can also strengthen the linear programming relaxation (LP relaxation) of F0 if we further add the following set of inequalities (\ref{eqn:f21}).

\begin{equation}\label{eqn:f21}
	\displaystyle\sum_{k \in M_{ij}} x_{ij}^k + \displaystyle\sum_{k \in M_{ij}} x_{ji}^k \leq 1\;, \;\;\forall (i,j) \in A_0 \mbox{ with } i < j \mbox{ and } |M_{ij}| \geq 1
\end{equation}

These are set packing constraints with cardinality 2 that inhibit a cycle among any pair of nodes $\{i,j\}$ if the two arcs $(i,j)$ and $(j,i)$ belong to the original graph. These inequalities integrate the class of set packing constraints, being the particular case involving just a pair of nodes (see, e.g., \cite{Gou1995}).

These two sets of inequalities are redundant in the original version of F0, but can strengthen the associated LP relaxation as shown in Section \ref{Sec4:Tests}. We will denote by F1 the original formulation F0 augmented with constraints (\ref{eqn:f20}). When we further include inequalities (\ref{eqn:f21}) in F1, the model becomes F2.


\section{Computational tests} \label{Sec4:Tests}

The tests were carried out on a laptop with the following specifications: an Intel Core i5-12500H processor, 8 GB of RAM, and the Windows 11 operating system. The experiments were developed in Python, using the Gurobi solver, version 12.0.2. To obtain the geographical coordinates, as well as the distances and travel times between different clients, a Microsoft Azure API was used.

In all test cases, the default parameters settings of Gurobi were maintained when running the branch-and-bound. In addition, the maximum time was set to 3600 seconds and the branch-and-bound duality gap was set to 0.0001, ensuring consistent stopping criteria across all instances.

In Subsection \ref{Sec4.1:Tests1}, we start discussing the computational performance of the formulation, considering the additional constraints that strengthen the associated linear programming relaxation, and also discuss the influence of the main parameters that characterize the problem. Then, in Subsection \ref{Sec4.2:Tests2}, the real applicability of the problem and the model to support decision making is discussed, comparing the solutions produced by the formulation with those built by the company.

\subsection{Computational performance of the formulation} \label{Sec4.1:Tests1}

Initially, tests were carried out to evaluate the performance of the proposed formulation. For this purpose, widely used benchmark instances from the literature were employed, namely the Solomon instances (1987) (see, e.g., \cite{Dumasetal1991}), designed for the Vehicle Routing Problem with Time Windows (VRPTW). These instances are divided into three groups: C (clustered customers), R (randomly distributed customers) and RC (a combination of both). The suffix 1 indicates tighter time windows and lower vehicle capacity, while suffix 2 corresponds to wider time windows and higher capacity. The planning horizon also varies, being shorter in the instances with suffix 1.
However, these instances do not take into account specific characteristics of the context under study, such as company working hours, maximum daily working time per employee, and lunch breaks. In the present work, vehicles are interpreted as employees, and capacity constraints (in the sense of volume or weight) are not considered. On the other hand, the lunch break is explicitly included in the model, which required adapting the original instances.

In all cases, 25 employees were considered, corresponding to the number of vehicles available in the Solomon cases. The lunch break interval was proportionally adjusted, considering the high variability of the planning horizon, which ranges from 230 to 3390 time units and the demanding time windows. To ensure feasibility, the lunch break was placed approximately halfway through the planning horizon.


In total, the 56 Solomon instances were tested, with only 4 reaching the optimal solution or a feasible solution within the given time limit. These adaptations made it possible to assess the robustness of the proposed formulation in a more realistic context, while respecting the operational constraints of the problem under study.

Table~\ref{tab:t1} shows the results obtained for the adapted Solomon instances that found, at least, one feasible solution.

\begin{table}[h]
	\caption{Performance results for the adapted Solomon instances.}
	\label{tab:t1}
	\centering
	{\footnotesize
		\begin{tabular}{l c}
			\hline
			\textbf{Instance} & \textbf{OPT/GAP} \\
			\hline
			\textbf{C1} & \\
			\quad C105 & 70.66\% \\
			\quad C106 & 64.95\% \\
			\hline
			\textbf{C2} & \\
			\quad C201 & 0.76\% \\
			\quad C205 & OPT \\
			\hline
		\end{tabular}
	}
\end{table}

Considering the differences between the problem addressed in this work and those in the literature that also resort to the Solomon instances, a direct comparison of the results is limited. Comparisons were mainly made with the study by \cite{KarNaw2022}, which, although similar in the use of time windows and distance minimization, does not consider lunch breaks and uses modified instances, making a fair comparison difficult. Moreover, their tests were limited to instances with up to 50 customers, whereas this work uses 100 customers.

Other studies, such as \cite{TanKLeeLZhu2001}, \cite{RocTai1995} and \cite{OmRosHans2006} apply heuristic approaches and report results in terms of total distance and number of vehicles, without reporting gaps, and generally use fewer vehicles than the 25 fixed employees considered in this work.

For the adapted Solomon instances, the proposed methodology achieved competitive results in terms of total distance, despite using more employees. For example, in instances C105 and C106, the obtained distances were significantly lower than those reported in the literature, although with more vehicles. Only 4 of the 56 Solomon instances produced feasible solutions within the studied context, leading the remaining tests in this section to be conducted using real data provided by the company.

The real instances are denoted by r-n-m, with the $n$ clients geographically located in the municipality of Marco de Canaveses, in Portugal. The set of clients in each instance are taken randomly among the entire clients of the company. Most of these clients are located within the city, while others are located outside, including other cities in other municipalities. The $m$ collaborators are characterized by the subset of clients they cover and their own labor time. 

To assess the impact of redundant constraints (\ref{eqn:f20}) and (\ref{eqn:f21}) in the proposed integer formulation, computational experiments were carried out with and without their inclusion, using instances of varying sizes. The objective is to determine whether these constraints significantly affect the linear relaxation of the model and, consequently, the performance of the Branch-and-Bound algorithm used for its resolution.
Table~\ref{tab:t2} presents the results obtained for each of the three formulations, comparing linear relaxation bounds and overall Branch-and-Bound performance. The tests were conducted on the previously described real instances, evaluating the inclusion of constraint (\ref{eqn:f20}), constraint (\ref{eqn:f21}), or both.

\begin{table}[h]
	\caption{Comparative Analysis and Performance of Formulations F0, F1, and F2}
	\label{tab:t2}
	\centering
	{\footnotesize
		\begin{tabular}{l c c c c c c c c}
			\hline
			Instance & $n$ & $m$ & F0 OPT/GAP & F0 time & F1 OPT/GAP & F1 time & F2 OPT/GAP & F2 time \\
			\hline
			r-10-2 & 10 & 2 & OPT & 0.28 & OPT & 0.28 & OPT & 0.33 \\
			r-12-2 & 12 & 2 & OPT & 100.84 & OPT & 205.13 & OPT & 87.05 \\
			r-15-2 & 15 & 2 & 45.66\% & 3600 & 44.88\% & 3600 & 42.68\% & 3600 \\
			r-20-3 & 20 & 3 & 49.60\% & 3600 & 49.55\% & 3600 & 52.62\% & 3600 \\
			r-25-3 & 25 & 3 & 54.42\% & 3600 & 54.31\% & 3600 & 54.34\% & 3600 \\
			r-30-4 & 30 & 4 & 51.04\% & 3600 & 50.86\% & 3600 & 51.84\% & 3600 \\
			r-40-7 & 40 & 7 & 44.19\% & 3600 & 43.81\% & 3600 & 41.32\% & 3600 \\
			r-50-7 & 50 & 7 & 68.18\% & 3600 & 66.88\% & 3600 & 64.53\% & 3600 \\
			r-60-9 & 60 & 9 & 64.59\% & 3600 & --------- & 3600 & 62.06\% & 3600 \\
			r-70-12 & 70 & 12 & 63.64\% & 3600 & 65.81\% & 3600 & 62.93\% & 3600 \\
			r-80-18 & 80 & 18 & --------- & 3600 & --------- & 3600 & --------- & 3600 \\
			r-90-22 & 90 & 22 & 44.73\% & 3600 & 44.28\% & 3600 & 21.36\% & 3600 \\
			r-96-24 & 96 & 24 & 23.21\% & 3600 & 20.17\% & 3600 & 24.44\% & 3600 \\
			\hline
		\end{tabular}
	}
\end{table}

The comparison between formulations F0, F1, and F2 shows a progressive improvement in results as additional constraints are introduced, with a particularly significant contribution from constraints (\ref{eqn:f21}) that characterize formulation F2. This improvement is evident in both the linear relaxation bounds and in the overall performance of the Branch-and-Bound algorithm, especially for larger instances.

Although adding constraints increases computational effort, this cost is often offset by the strengthening of the linear relaxation, which can enhance the efficiency of the Branch-and-Bound process. However, this trade-off is not always straightforward, as a more complex formulation may also negatively affect performance in certain cases.

In general, formulation F2 showed to be the most effective and will therefore be adopted in the subsequent analyses. It is also worth noting that the reported gap refers to the relative difference between the lower and upper bounds, which may not represent the true distance to the optimal solution. Based on practical analysis, geographic inspection of the solutions, and expert knowledge from the company, it is believed that the feasible solutions (upper bounds) obtained are close to the true optimum, with the gap largely attributed to weak lower bounds.

\vspace{1em}
The performance of formulation F2 is also analyzed with respect to variations in the number of available employees, aiming to assess the impact of this parameter on solution quality and computational effort. Additionally, two assignment strategies are compared: in version F2-a, each employee can only serve a subset of customers (as initially proposed in this work), while in version F2-b, all employees can serve any customer, as in traditional vehicle routing problems. In both cases, identical time windows are considered. The results are presented in Table~\ref{tab:t3}.

\begin{table}[h]
	\caption{Comparison and performance evaluation of Formulation F2, concerning the sets of customers covered by each employee}
	\label{tab:t3}
	\centering
	{\footnotesize
		\begin{tabular}{l c c c c c c}
			\hline
			Instance & $n$ & $m$ & F2-a OPT/GAP & F2-a time & F2-b OPT/GAP & F2-b time \\
			\hline
			r-10-2 & 10 & 2 & OPT & 0.27 & OPT & 18.35 \\
			r-10-4 & 10 & 4 & OPT & 0.11 & OPT & 0.10 \\
			r-10-6 & 10 & 6 & OPT & 0.23 & OPT & 0.18 \\
			r-30-4 & 30 & 4 & 32.65\% & 3600 & --------- & 3600 \\
			r-30-8 & 30 & 8 & OPT & 14.80 & 9.11\% & 3600 \\
			r-30-14 & 30 & 14 & OPT & 11.10 & OPT & 24.24 \\
			r-60-9 & 60 & 9 & 40.60\% & 3600 & --------- & 3600 \\
			r-60-18 & 60 & 18 & 25.14\% & 3600 & 28.62\% & 3600 \\
			r-60-28 & 60 & 28 & 29.21\% & 3600 & OPT & 1258.61 \\
			r-80-18 & 80 & 18 & 25.82\% & 3600 & --------- & 3600 \\
			r-80-36 & 80 & 36 & 26.92\% & 3600 & OPT & 1547.52 \\
			r-90-22 & 90 & 22 & 23.68\% & 3600 & --------- & 3600 \\
			r-90-44 & 90 & 44 & 0.97\% & 3600 & 75.27\% & 3600 \\
			\hline
		\end{tabular}
	}
\end{table}

The value of $m$ (number of employees) was initially defined as the smallest value for which a feasible solution could be found within the time limit. From this point on, $m$ was gradually doubled in order to expand the solution space and analyze the impact of increasing the number of employees on the quality of the solution and the resolution time.

The comparison between F2-a and F2-b reveals relevant performance differences. In small problems (n = 10), both versions reach optimality, although F2-b performs better in most cases. For n = 30, F2-a is shown to be more efficient, with shorter execution times and greater consistency in achieving the optimal solution. In larger instances, however, F2-b tends to be more effective, benefiting from greater routing flexibility, which enhances the exploration of the solution space.

Overall, formulation F2-a performs better when the number of employees is small, while F2-b becomes more advantageous as $m$ increases, in general.


An additional analysis assessed the impact of time windows on the computational performance of formulation F2, testing three configurations: without time windows (F2-w), with regular time windows (F2-r) and with tight time windows (F2-t).

\begin{itemize}
	\item F2-w: service allowed at any time, just observing the company and employees' operating periods.
	\item F2-r: regular time windows, typically morning or afternoon shifts, with a minimum duration of three hours, offering great flexibility.
	\item F2-t: tight time windows ranging from 15 minutes to 2 hours, often 30 minutes.
\end{itemize}

All formulations include tight time windows for fictitious clients (lunch break, [12:30 - 12:45]) and clients without scheduled time windows, in line with the company's usual operation process. Note that the time windows restrict the start of the visit, not its completion.

\begin{table}[h]
	\caption{Comparison and performance evaluation of Formulation F2, considering versions F2-w, F2-r, and F2-t for time windows settings}
	\label{tab:t4}
	\centering
	{\footnotesize
		\begin{tabular}{l c c c c c c c c}
			\hline
			Instance & $n$ & $m$ & \multicolumn{2}{c}{F2-w} & \multicolumn{2}{c}{F2-r} & \multicolumn{2}{c}{F2-t} \\
			& & & OPT/GAP & time & OPT/GAP & time & OPT/GAP & time \\
			\hline
			r-10-2 & 10 & 2 & OPT & 0.33 & OPT & 0.23 & OPT & 0.13 \\
			r-12-2 & 12 & 2 & OPT & 87.05 & OPT & 24.05 & OPT & 0.75 \\
			r-15-2 & 15 & 2 & 42.65\% & 3600 & 23.95\% & 3600 & OPT & 597.02 \\
			r-20-3 & 20 & 3 & 52.62\% & 3600 & 50.03\% & 3600 & 32.63\% & 3600 \\
			r-25-3 & 25 & 3 & 54.34\% & 3600 & 43.57\% & 3600 & --------- & 3600 \\
			r-30-4 & 30 & 4 & 51.84\% & 3600 & 50.61\% & 3600 & 25.81\% & 3600 \\
			r-40-7 & 40 & 7 & 41.32\% & 3600 & 23.59\% & 3600 & 15.92\% & 3600 \\
			r-50-7 & 50 & 7 & 64.53\% & 3600 & 31.59\% & 3600 & 22.56\% & 3600 \\
			r-60-9 & 60 & 9 & 62.06\% & 3600 & 49.76\% & 3600 & 35.94\% & 3600 \\
			r-70-12 & 70 & 12 & 62.93\% & 3600 & 56.47\% & 3600 & 43.18\% & 3600 \\
			r-80-18 & 80 & 18 & --------- & 3600 & 45.89\% & 3600 & 30.40\% & 3600 \\
			r-90-22 & 90 & 22 & 21.36\% & 3600 & 21.50\% & 3600 & --------- & 3600 \\
			r-96-24 & 96 & 24 & 24.44\% & 3600 & 52.01\% & 3600 & --------- & 3600 \\
			\hline
		\end{tabular}
	}
\end{table}

Modeling time windows in vehicle routing problems increases the formulation size, raising computational effort, but simultaneously restricts the feasible solution space, which can reduce resolution time. For small and medium-sized instances, especially with tight time windows, better performance is observed due to this solution space reduction (\cite{GhoGha2010}). However, in larger instances, overly restrictive time windows hinder resolution, sometimes resulting in no feasible solutions. Thus, time windows improve model performance in smaller problems but may impair it at larger scales due to the increased number of constraints.

\subsection{Testing the real applicability of the problem} \label{Sec4.2:Tests2}

Next, we analyze case studies applied within the organizational context. When describing the routes, $\alpha$ represents the lunch break (fictitious client). 

We start with a small instance involving 10 clients and 2 collaborators ($n$ = 10, $m$ = 2). The locations of the condominiums (clients) are in the northern region of Portugal, covering the districts of Porto and Viseu. Nine clients are located in Porto (eight in Marco de Canaveses and one in Penafiel) and one in Cinf\~{a}es (Viseu). Seven clients are in urban areas and three in rural areas. The time windows are tight (maximum two hours), with overlapping windows for clients 7 and 90, preventing visits on the same route. Collaborators A and B have working periods of 8 and 6 hours, respectively. Considering the specificity of the requested services and the skills of the employees, clients \{20, 32, 80\} can be served only by employee A, clients \{7, 25, 41, 45\} can be served only by employee B, and the remaining clients (\{8, 83, 90\}) can be served by either employee. Figure \ref{fig:f1a} in the Appendix shows the locations of the clients in the map.

Formulation F2 found the optimum solution in 0.36 seconds. This solution and the one built by the company (manual routes) are reported in Table \ref{tab:t5}.

\begin{table}[h]
	\centering
	{\footnotesize
		\renewcommand{\arraystretch}{1.3}
		\caption{Comparison between the manual routes and the routes obtained from the formulation}
		\label{tab:t5}
		\begin{tabular}{l c c}
			\hline
			Routes & A & B \\
			\hline
			Manual      & 90 20 80 83 32 & 8 25 41 $\alpha$ 7 45 \\
			Formulation & 90 8 32 80 20  & 25 83 $\alpha$ 7 41 45 \\
			\hline
		\end{tabular}
	}
\end{table}

The two solutions can be seen in Figure \ref{fig:f1b} in the Appendix, for employees A and B.

There are minor differences between the routes. In the manual routes, employee A visits client 83 instead of client 8. However, in both routes, client 90 is visited by employee A, although it could be served by either of them. This situation was expected, as the time window for client 90 coincides with that of client 7, as previously mentioned. Still within both solutions, the route assigned to employee A is entirely scheduled for the afternoon period.
In the case of employee B, the route covers both morning and afternoon, which requires the inclusion of a fictitious client for lunch break.
In the manually defined routes, employee B must visit clients 8, 25, and 4 in the morning. In the optimized solution, the same employee is required to visit only clients 25 and 83 in the morning, leaving the remaining visits for the afternoon.
Overall, across the two routes for both employees, the difference in total distance traveled is approximately 2 km, corresponding to only a 2 minutes difference in time. This difference, although small, occurs mainly in employee's A route.

\vspace{1em}

Another tested instance involved 12 clients ($n$ = 12), also located in the northern region of Portugal, mainly in the municipality of Marco de Canaveses, with one client in Castelo de Paiva (Aveiro district). It includes 11 clients in urban areas and 1 in the outskirts. For this scale, the company uses 2 employees ($m$ = 2), each with eight-hour work schedules.
Considering the specific nature of the services and the skills of each employee, some clients can only be visited by employee A (\{12, 26, 65\}) and others only by employee B (\{24, 39\}), while the remaining clients (\{9, 16, 35, 55, 69, 88, 96\}) can be served by either employee. Time windows vary between regular and tight. The time windows for clients 69 and 88 coincide (16:00-16:30), making it impossible for them to be visited by the same employee.
The optimal solution was obtained after 1.59 seconds. Both solutions are represented in Table \ref{tab:t6}.

\begin{table}[h]
	\centering
	{\footnotesize
		\renewcommand{\arraystretch}{1.3}
		\caption{Comparison between the manual routes and the routes obtained from the formulation}
		\label{tab:t6}
		\begin{tabular}{l c c}
			\hline
			Routes & A & B \\
			\hline
			Manual      & 65 9 $\alpha$ 12 88 35 26 & 16 24 39 55 $\alpha$ 69 96 \\
			Formulation & 12 69 26 65              & 9 55 24 $\alpha$ 16 39 88 96 35 \\
			\hline
		\end{tabular}
	}
\end{table}

This case reveals an unexpected result: the manual route has a shorter distance (around 45 km) compared to the route obtained through the formulation (around 50 km). This difference is explained by the fact that the manual route includes a lunch break for both employees, whereas the formulation only assigns this break to employee B.
Since the objective function reflects the practical needs of the company, the lunch break should only be included when strictly necessary. Thus, although the formulation does not minimize the distance as efficiently, it respects the preference to concentrate on the workday either in the morning or in the afternoon whenever possible, which justifies the observed difference. Overall, the manual route has an advantage of about 5 km and 10 minutes. However, this version does not account for the penalty associated with including the lunch break, which is incorporated into the objective function.
If this penalty is removed, allowing the lunch break to be considered flexibly, the formulation produces a more efficient route, with 2 km less than the manual route. Nevertheless, this solution forces employee A to work in both the morning and afternoon, reducing their availability for other tasks. Considering that, in this sector, many employees are subcontracted and not exclusively dedicated to a single company, the penalty for including the lunch break in the objective function is therefore justified. In effect, a route starting in the middle of the morning and ending in the middle of the afternoon could be rather damaging for the productivity of these employees.

\vspace{1em}

In the case with 15 clients ($n$ = 15), still located in the northern region of Portugal, distributed across the districts of Porto and Aveiro. Only one client belongs to the district of Aveiro, situated in Castelo de Paiva (client 46); the remaining clients are located in the municipality of Marco de Canaveses, except for client 56, who is based in the municipality of Penafiel. The set includes 11 clients in urban areas and 4 in non-urban areas.

For this size of the problem, the company also assigns two employees, each working an eight-hour shift. Due to the specific requirements of the services and the competencies of each employee, some clients are exclusive to a single employee: clients \{12, 17\} can only be served by employee A, while clients \{70, 88\} are exclusive to employee B. The remaining clients (\{2, 9, 20, 40, 41, 42, 46, 49, 56, 58, 92\}) can be served by either employee. Figure \ref{fig:f2a} in the Appendix shows the locations of the clients in the map.

The time windows include tight 30 minute intervals and broader morning-long intervals. In particular, two clients share the same time window: clients 2 and 46, both scheduled between 10:30 and 11:00.

The optimal solution was obtained in 1073.18 seconds. Both solutions are represented in Table \ref{tab:t7}.

\begin{table}[h]
	\centering
	{\footnotesize
		\renewcommand{\arraystretch}{1.3}
		\caption{Comparison between the manual routes and the routes obtained from the formulation}
		\label{tab:t7}
		\begin{tabular}{l c c}
			\hline
			Routes & A & B \\
			\hline
			Manual      & 12 2 56 $\alpha$ 17 20 41 40 42 & 58 46 88 $\alpha$ 49 9 92 70 \\
			Formulation & 49 12 46 $\alpha$ 56 17 20 42 40 41 & 88 58 2 $\alpha$ 70 92 9 \\
			\hline
		\end{tabular}
	}
\end{table}

The routes of both employees differ significantly, and in both cases the routes cover both the morning and the afternoon. The two solutions can be seen in Figure \ref{fig:f2b} in the Appendix, for employees A and B. In the figure, the route of employee A appears to be a closed loop. In fact, the route is open, being broken in the area with the highest customer density (difficult to see in the image). The reason why the route forms an almost complete loop is due to customers' time windows, with customers who are very close together having time windows that are very far apart.

In the figure, the route of employee A seems a closed cycle, 

There is a considerable difference in the total distance between the manual routes and those obtained through the formulation, approximately 39 km. This discrepancy is mainly due to the inclusion of client 46 in the employee B manual route, who has a restricted time window in the mid-morning (10:30-11:00), making the route longer. To meet this window, the sequence of visits in the manual route requires detours that increase the total distance.
Additionally, since clients 2 and 46 have overlapping time windows, they must be served by different employees. In the construction of employee B manual route, including client 46 proved suboptimal, an inefficiency that only becomes apparent when compared with the solution generated by the formulation.

\vspace{1em}

The following table (Table ~\ref{tab:t8}) presents the practical applicability of the solutions obtained, involving instances with a larger number of clients. No scenario guaranties the optimal solution, but in some cases the upper bound may be close to the optimal value of the problem. For these cases, the company does not produce manual results.

\begin{table}[h]
	\centering
	{\footnotesize
		\renewcommand{\arraystretch}{1.3}
		\caption{Instances with a larger number of clients (20 and 96 clients)}
		\label{tab:t8}
		\begin{tabular}{l c c c c}
			\hline
			Instance & $n$ & $m$ & GAP & time \\
			\hline
			r-20-3  & 20  & 3  & 48.60\% & 3600 \\
			r-96-24 & 96  & 24 & 51.65\% & 3600 \\
			\hline
		\end{tabular}
	}
\end{table}

In the tests with 20 clients and 3 employees, the routes were consistent with time constraints and time windows, respecting client exclusivity and the need to focus on the workday in the morning or afternoon whenever possible. For example, the employee's A route covers all visits in the afternoon, ensuring that a lunch break is included only when necessary. The routes of employees B and C also show logical sequences, taking into account time windows and the geographical distribution of clients, ensuring efficiency even in complex situations.

For the dataset with 96 clients and 24 employees, despite the gap being even higher, the routes remain coherent and well-structured. Most routes are completed entirely in the morning or afternoon, and when necessary, a lunch break is included to accommodate clients with longer time windows or services that require more time. The sequence of visits is geographically logical and consistent.

In summary, even with high gaps, the routes demonstrate a high level of organization, respect the time windows and each employee's requirements, indicating that the solutions are practically applicable and efficient, although it cannot be said with certainty that they are optimal.

\vspace{1em}

A case with 40 clients ($n$ = 40) and 7 employees ($m$ = 7) was also analyzed, considering three scenarios of time windows: no time windows, regular time windows (wide intervals, ($\ge 3$ hours) and tight time windows (short intervals, $\le 1.5$ hours).
In the scenario without time windows, route management is more flexible, resulting in fewer employees with full-day routes and a reduced need for lunch breaks. In scenarios with time windows, the requirement to meet specific schedules increases planning rigidity, the need for more breaks and, in general, detours that increase the total distance travel.
Interestingly, the scenario with regular time windows shows a slightly lower total distance than the scenario without time windows. This is mainly due to the fact that neither solution reaches the optimum and differences can be hidden within the duality gaps. Yet, we can also detach that the more frequent inclusion of the fictitious client to comply with the time windows, which, despite increasing breaks, allows for a more optimized sequence of visits and slightly improves distance efficiency. Nevertheless, the company still prioritizes routes that are completed entirely in the morning or afternoon.

\section{Conclusions} \label{Sec5:Conclusions}

This paper proposes a mathematical formulation for planning employees' routes in a condominium management company, subject to operational constraints that include restrictions on the routes and labor issues, focused on the minimization of total distance traveled and penalizing long duration routes that force lunch breaks. The formulation can also be adapted to other companies that also provide field service operations. The approach proved effective by automatically creating coherent and adequate routes, making better use of available time and reducing unnecessary travel, confirming its applicability to the company's operational reality.

The computational tests showed the efficiency and effectivity of the proposed formulation for moderate sized instances. It also helped to analyze the model performance and the quality of the solutions under different parameters' scenarios, from an applied perspective.

The solutions brought by the model also revealed important advantages compared to those usually built by the company. In this way, the model stands for its ability to automatically provide good solutions to support the decision-making process involved in daily route planning for field service operations in a condominium company.

Several directions are suggested for future work: expanding the model to allow multiple employees to serve the same client simultaneously; allowing non-coverage of all customers, with a penalty; integrating daily planning into a weekly horizon; or testing alternative objective functions. The implementation of approximate methods would also allow the model to be applied to larger-scale companies.

In summary, the proposed formulation contributes to the continuous improvement of logistical processes, providing effective and efficient solutions while establishing a solid foundation for future extensions and applications.

\section*{Acknowledgements}
Pedro Martins acknowledges support from the FCT - Funda\c{c}\~{a}o para a Ci\^{e}ncia e a Tecnologia, I.P., through national funds, under the project UID/04561/2025. The authors also acknowledge support by Coimbra Business School - ISCAC, Polytechnic University of Coimbra.

\section*{Appendix} \label{App1}
				
\begin{figure}[h]  
	\centering
	\includegraphics[width=\textwidth]{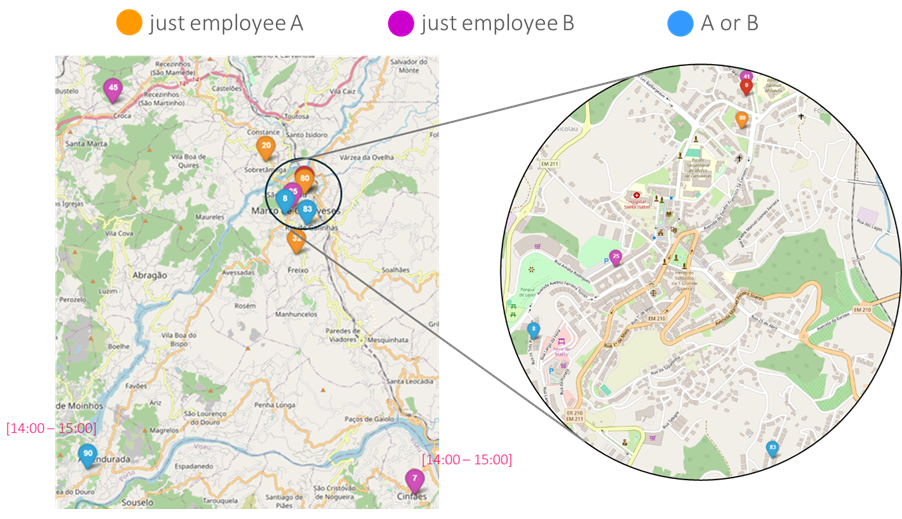}
	\caption{Location of the clients in the map for the $n=10$ clients instance.}
	\label{fig:f1a}
\end{figure}

\begin{figure}[h]  
	\centering
	\includegraphics[width=\textwidth]{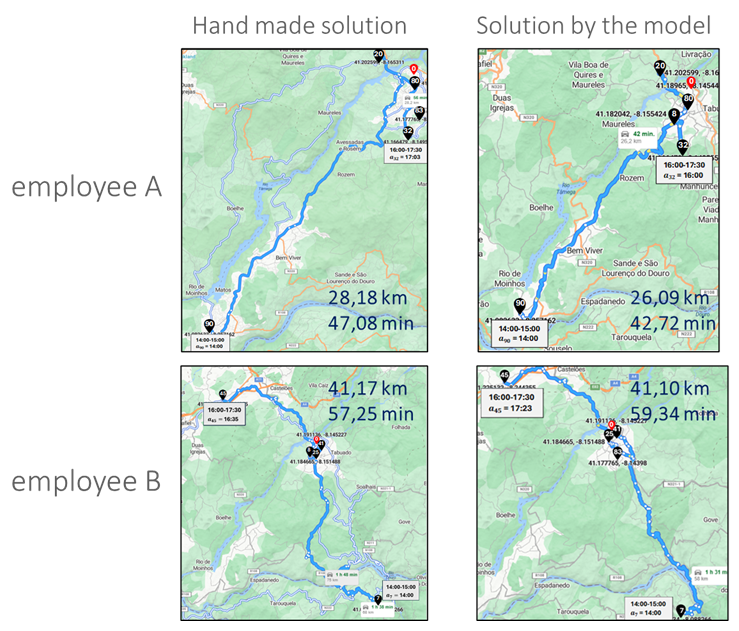}
	\caption{Hand made solution built by the company and the one obtained using the formulation, for the $n=10$ clients instance.}
	\label{fig:f1b}
\end{figure}

Real instance with $n=10$ clients and $m=2$ employees considered in Subsection \ref{Sec4.2:Tests2}. Figures \ref{fig:f1a} and \ref{fig:f1b} show the locations of the clients in the map and the solution built by the company and the one obtained using the formulation.
				
\begin{figure}[h]  
	\centering
	\includegraphics[width=\textwidth]{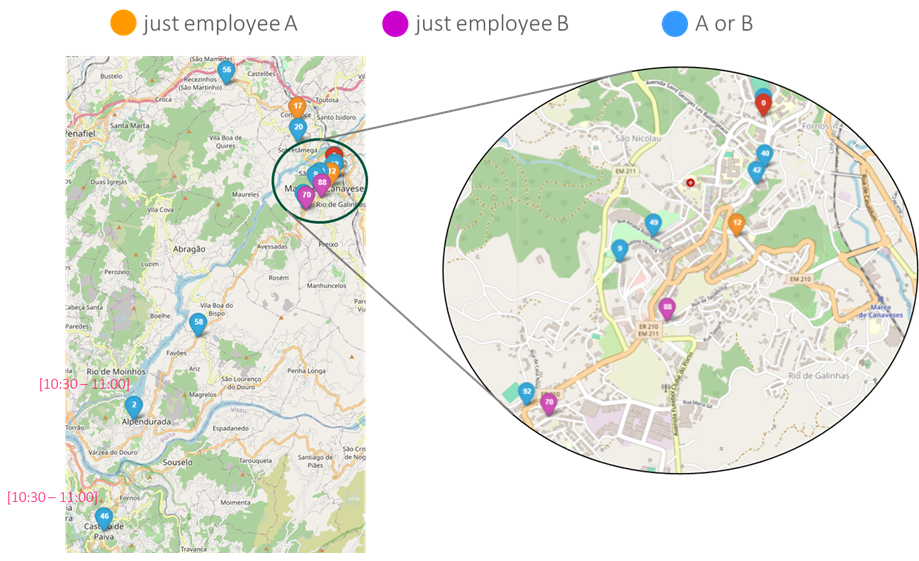}
	\caption{Location of the clients in the map for the $n=10$ clients instance.}
	\label{fig:f2a}
\end{figure}

\begin{figure}[h]  
	\centering
	\includegraphics[width=\textwidth]{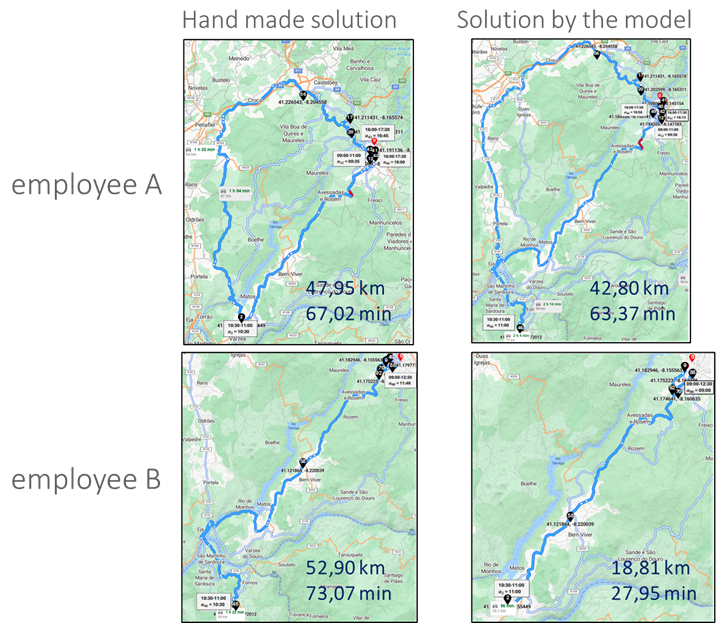}
	\caption{Hand made solution built by the company and the one obtained using the formulation, for the $n=10$ clients instance.}
	\label{fig:f2b}
\end{figure}

Real instance with $n=15$ clients and $m=2$ employees considered in Subsection \ref{Sec4.2:Tests2}. Figures \ref{fig:f2a} and \ref{fig:f2b} show the locations of the clients in the map and the solution built by the company and the one obtained using the formulation.
				
\end{document}